\documentclass{jfm}
\usepackage{graphicx}
\usepackage{newtxtext}
\usepackage{newtxmath}
\usepackage{natbib}
\usepackage{hyperref}
\hypersetup{
    colorlinks = true,
    urlcolor   = blue,
    citecolor  = black,
}

\newcommand{\RomanNumeralCaps}[1]
\linenumbers
\usepackage{graphicx}
\usepackage{epstopdf, epsfig}
\usepackage{chemformula}
\usepackage{siunitx}
\usepackage{booktabs}

\shorttitle{Pore scale reaction}
\shortauthor{S. Izumoto, G. Rousseau, T. Le Borgne and J. Heyman}

\title{Impact of pore-scale chaotic mixing on Darcy-scale reaction rates}

\author{Satoshi Izumoto\aff{1,2}
\corresp{\email{satoshi.izumoto@hotmail.co.jp}},
  Gauthier Rousseau\aff{1}
  Tanguy Le Borgne\aff{1}
 \and Joris Heyman\aff{1}}

\affiliation{\aff{1}Univ. Rennes 1, CNRS, Géosciences Rennes, UMR 6118, 35000 Rennes, France\aff{2}Institute de Physique de Rennes, Univ. Rennes 1, CNRS, Unite Mixte de Recherche 6118, Rennes, France}

\begin{document}

\maketitle

\begin{abstract}
Prediction of reactive transport in porous media remains challenging when pore scale incomplete mixing is at play. Previous experimental studies investigated chemical reactions in porous media by visualizing reaction product or reactants mostly in uniform flow. However, the local reaction rate, which is necessary to infer mechanisms of reaction in pore space, could not be obtained without considering transport of reaction products and reactants. Thus, the interpretation remained elusive. We visualized the reaction rate field using chemiluminescnece within index-matched 3D porous media under zero acceleration and constant acceleration flow fields to investigate how pore scale chaotic mixing and Darcy scale fluid acceleration rectify reactive transport. We found that the reaction rate kept increasing from upstream to downstream in constant acceleration field, whereas it increased only at the upstream zone in zero acceleration field. The ratio of dispersion rate and size of the mixing interface determined such an effect of acceleration. Moreover, the experimental results showed stronger dependency of reaction rate on velocity compared to the numerical simulations that assume complete mixing in pore space. To explain this, we suggested the mechanistic model that includes the pore scale folding of lamellae due to chaotic mixing and the pore scale concentration gradients against compression. Such a pore scale mechanism was consistent with the experimentally observed change in reaction rate over the space. These results give new insights on underlying mechanisms of reactive transport in porous media.
\end{abstract}

\begin{keywords}
Authors should not enter keywords on the manuscript, as these must be chosen by the author during the online submission process and will then be added during the typesetting process (see http://journals.cambridge.org/data/\linebreak[3]relatedlink/jfm-\linebreak[3]keywords.pdf for the full list)
\end{keywords}

\maketitle

%\tableofcontents

\section{Introduction}
Chemical reactions in porous media alter the transport, the transformation and the degradation of chemical and biological substances in subsurface environments, such as soils and aquifers \citep{Chapelle2001}. The spatial and temporal heterogeneity of fluid flows in natural porous systems have important consequences for reactive processes, such as localization of reaction in hot-spots \citep{Bochet2020,McClain2003} or hot-moments \citep{Briggs2014,Gu2012}, and long-lasting reaction at geological time scale \citep{Hilley2008}. In turn, reactive transport plays a key role in practical applications, including remediation of contaminated ground water \citep{Fu2014} and geological carbon sequestration \citep{Zoback2012}. \\
Many classical experiments investigated the reactive transport in porous media using column setups that allow point measurements of concentrations of reactive tracers \citep{Raje2000,Valocchi2019b}. Only a few studies visualized the concentration fields of reactive tracers in 3D porous media, which have rich information of the impact of incomplete mixing on reaction. \citet{Gramling2002} used refractive index matching technique to visualize the concentration field of reaction product while a solution of one reactant displaces the other solution with the other reactant. This study highlighted that continuum scale approximation cannot be applied for reactive transport when pore scale incomplete mixing plays major role. \citet{Edery2015} also applied refractive index matching technique and visualized the evolution of the concentration field of reactant under acid-base reaction, where a solution with lower pH was injected from a point in a macroscopically uniform flow field of solution with higher pH. This study highlights the usefulness of continuous time random walk formulation to model the reactive transport including pore scale incomplete mixing (termed as small-scale fluctuations in the paper), where the parameters were determined from conservative tracer experiments and batch experiment for chemical reactions, and no fitting parameter was required. More recently, \citet{Markale2021} used magnetic resonance imaging to visualize the reaction product of invading reactive front in pore scale, where the study observed persistent reaction behind the reactive front due to incomplete mixing.\\
These studies pointed out how important pore scale incomplete mixing is to understand reactive transport in porous media, and suggested how to model the reactive transport. However, the interpretation of the results was limited because they have transient mixing front and transport of reaction products or reactants. Particularly, it is known that 3D porous media induces stretching and folding of the fluid element, called chaotic mixing, in pore space \citep{Metcalfe2022}. A previous study showed experimentally that chaotic mixing induced exponential elongation of the mixing interface of the conservative tracer \citep{Heyman2020}, but it has been not yet investigated how such a strong mixing influences the reactive transport. To infer pore scale mechanisms that control reactions, it is necessary to visualize the reaction rate field instead of reaction products and reactants.\\
Some previous studies further investigated the reaction under Darcy scale heterogeneous flow field such as flow focusing \citep{Rolle2009} and helical flow \citep{Ye2020}, and reported that such heterogeneity enhances reaction. The enhancement of reaction is important for the application of reactive transport in engineering problem, but these studies stay in specific case studies for a specific flow field. More general understanding of how Darcy scale flow acceleration and deceleration impact reactive transport is necessary to understand reactive transport in heterogeneous flow fields.\\
Here, we experimentally visualized stationary reaction rate field in 3D porous media when the reaction rate is fast enough to limit the reaction zone within a few pore spaces, where pore scale incomplete mixing is particularly important. We chose two types of flows to investigate how Darcy scale fluid acceleration impact the reactive transport; co-flow and saddle flow. The former has zero compression rate whereas the latter has constant compression rate everywhere (constant acceleration). By comparing to the results of numerical simulations that assume complete mixing in pore scale, we suggest that the chaotic mixing in pore scale enhances reaction by creating layers of reactants.

\section{Method}
\begin{figure}
    \centering
    \includegraphics{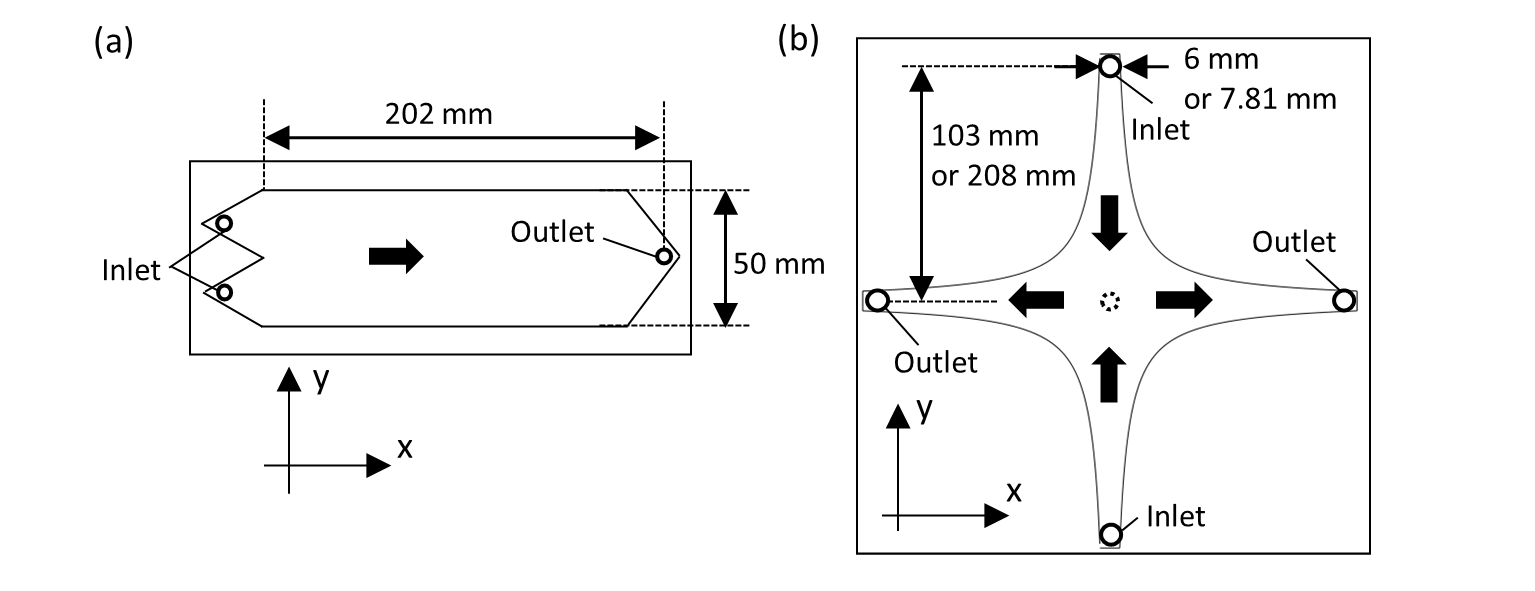}
    \caption{(a) Co-flow setup and (b) saddle-flow setup. The thick arrows indicate the flow direction}
    \label{figSetup}
\end{figure}

For the reactive transport experiments, we used luminol chemiluminescence, which is one of the most popular chemiluminescence reactions. This technique allows visualizing the reaction rate field instead of the reaction products or reactants. The chemiluminescence reaction involves a catalytic reaction of \ch{H2O2} with \ch{Co^{2+}} followed by an oxidation reaction of luminol with \ch{OH}$\cdot$ and \ch{O_2}$\cdot$ radicals \citep{Uchida2004}. This chain reaction can be written as:
\begin{equation}
\label{eq34}
    \ch{Luminol + H2O2 + 2 OH- ->[ Co^2+ ] 3 -Aminophthalatedianion + N2 + H2O + hv}
\end{equation}
Luminol is thus oxidized to 3-Aminophtalatedianion with an emmision of blue light ($\lambda =420-460$ nm). The reaction rate is proportional to the blue light intensity in the image. 
The mixing interface of this reaction can be simulated with the bimolecular second-order reaction 
\begin{equation}
    \ch{A + B -> C + \text{photon}}
    \label{eq29}
\end{equation}
where $A$ and $B$ are associated to the \ch{H2O2} and the luminol species respectively \citep{Matsumoto2015}. To induce the luminol reaction, we prepared two solutions. One was a mixture of 1 mM luminol, 7 mM \ch{NaOH} and 0.01 mM \ch{CoCl2} (termed as luminol solution), and the other was a mixture of 0.5 mM \ch{H2O2} and 3.9 mM \ch{NaCl} (termed as \ch{H2O2} solution). In our previous study, we have estimated the reaction constant $k$ as 2.56 \si{s^{-1}.mM^{-1}} by mixing the two solutions in a beaker and measuring the intensity of light over time. The luminol reaction continues for long time with very small constant reaction rate after finishing the fast bimolecular second order reaction. The light emitted from such long-lasting reaction were subtracted from the original images. The characteristic time scale for reaction is obtained by 
$\tau_R=1/kC_0=2.7$, where $C_0$ is the concentration of luminol.\\
We chose two types of typical flow fields; co-flow and saddle flow (Fig.\ref{figSetup}). The saddle flow ensures constant compression rate in the entire domain, whereas the co-flow ensures zero compression rate. The flow field is defined as: $(v_x,v_y)=(Const., 0)$ for co-flow, and $(v_x,v_y)=(\gamma x, -\gamma y)$ for saddle-flow, where $\gamma$ is the compression rate. For each flow field, we prepared Hele-Shaw cell (empty tank made of PMMA) and the porous media (tank made of PMMA filled with grains) as below. The co-flow cell includes two inlets in two separated triangular-shaped branches. By injecting two different solutions from each of the inlet, they start mixing at the edge of the separator, and they flow toward one outlet at the other side of the cell, which is placed 220 mm from the start of the mixing. For the Hele-Shaw cell, we set the width of the cell (50 mm) much larger than the height (2 mm) so that the boundary effects due to the side walls are negligible. In case of the porous media, we packed fluorinated ethylene propylene (FEP) grains (pellet shape, size: 2-3 mm) in the cell with larger height (12 mm). This enabled us to visualize the mixing interface in the porous medium because the refractive index of FEP (1.34) is close to that of water (1.33). The porosity of the packed FEP was calculated as 0.37 by measuring the weight of the packed FEP grains. In the saddle flow cell, there were four flow branches; two of them on opposite sides were for inlets and the two others for outlets. The shape of the walls follow $y = \pm a/x$, where $a$ is 303 $\si{mm}^2$ for the Hele-Shaw cell and 811 $\si{mm}^2$ for the cell for porous media. In Hele-Shaw cell, we set small height of the cell (2 mm) so that the effect of side wall on the flow is negligible at the middle of the cell. The distance between the inlet/outlet and the stagnation point was 103 mm. For porous medium, we used a larger cell (distance between the inlet/outlet and the stagnation point was 208 mm). The height was the same as the co-flow cell for the porous media (12 mm) and we packed the FEP grains.\\
The experimental protocol is as follows for all experimental configurations. We first filled the cell with deionized water. In case of porous media, we injected $\ch{CO2}$ gas before we inject the water so that the $\ch{CO2}$ gas dissolved into the water and there were no remaining bubbles. Then we injected the luminol solution and $\ch{H2O2}$ solution from two different injections at a certain injection rate. We imposed nine flow rates for each experimental configurations. For each flow rate, we wait for the front to be stabilized and then we took pictures by a mirrorless digital camera (14-bit, SONY alpha7s, SONY, Tokyo, Japan) with a macro lens (MACRO GOSS F2.8/90, SONY, Tokyo, Japan). The image resolution was 0.046 mm per pixel for all the cases. For porous media cases, we triplicated the experiments by repacking the FEP to obtain the global trend, which is independent of specific grain packing. The images were rescaled by the bit depth ($2^{14}-1$) to obtain the normalized reaction rate.\\
Each flow rate corresponds one Peclet number (Pe). For the co-flow, the Pe was calculated by $vL/D$, where $v$ is the velocity, $L$ is the characteristic size (2mm, size of the FEP grain and the height of the Hele-Shaw cell) and D is the diffusion coefficient ($1 \times 10^{-9}$ \si{m^2.s^{-1}}). For the saddle flow, Pe was $\gamma L^2/D$, where $\gamma$ is the compression rate estimated by $v_{inj}/L_{stag}$ with $v_{inj}$ the velocity at the injection and $L_{stag}$ the distance between the injection and the stagnation point. This resulted in Pe ranging from 179 to 3575 for co-flow and from 9 to 174 for saddle flow. \\
Since the luminol reaction continues for long time in small reaction rate constant, we subtracted the light emitted by such reaction in order to approximate luminol reaction as bimolecular second order reaction. We measured this weak light emission as follows. We first mixed the luminol solution and $\ch{H2O2}$ solution in a beaker, and then injected this solution in the Hele-Shaw cell and porous media. After more than 30 minutes, we took pictures. The intensity of these pictures correspond to the light emission from the long-lasting constant reaction. We subtracted these image intensities from the images taken in reactive transport experiments. Our previous study showed that this way of image processing allows us to interpret the image intensity as the light emission due to the bimolecular second order reaction, which is also checked by Hele-Shaw cell with saddle flow experiments in this study by comparing to the theoretical predictions.\\
In addition to the reactive tracer experiments, we performed conservative tracer experiments, where fluorescein sodium salt (12.5 \si{mg.L^{-1}}) and deionized water were injected instead of luminol solution and $\ch{H2O2}$ solution. We set the blue back-light panel behind the flow cells and green band-pass filter on the camera. We quantified the width of the mixing zone by fitting the error function for each concentration profile perpendicular to the mixing interface as $C = \left( 1  + \textrm{erf} \left(y/2w_c \right) \right)/2$, where C is the concentration of conservative tracer and $w_c$ is the mixing width. In case of saddle flow where compression rate is not zero, $w_c$ is called Batchelor scale representing the length scale that balances diffusion and compression. To avoid the artefact coming from the light scattering, we fitted only the concentration range above 0.5. More detailed experimental procedure are in our recent paper.\\
To help interpretation of the experimental results, we have run numerical simulations using open source CFD software OpenFOAM, which utilizes finite volume method. We consider the rectangular domain; for co-flow $x\in[0,150]$ mm with 300 meshes, $y\in[-25,25]$ mm with 400 meshes and for saddle-flow $x\in[-150,150]$ mm with 600 meshes, $y\in[-25,25]$ mm with 400 meshes. The flow field was $(v_x,v_y)=(\text{Const}., 0)$ for co-flow, and $(v_x,v_y)=(\gamma x, -\gamma y)$ for saddle flow, where $\gamma$ is the compression rate. The governing equation was;
\begin{equation}
\label{eq3}
    \frac{\partial C_A}{\partial t} = -v\cdot\nabla C_A+\nabla \cdot\left(D_{disp}\nabla C_A\right)-kC_AC_B
\end{equation}
where $D_{disp}$ is the dispersion tensor:
\begin{equation}
\label{eq4}
 D_{disp} = (D_m + \alpha |v|) I
\end{equation}
where $v$ is the velocity, $C_A$ and $C_B$ are concentrations of reactants, $D_m$ is the molecular diffusion set as $5\times10^{-10}$  $\si{m^2.s^{-1}}$ considering the tortuosity of porous media \citep{Sen1994,Scheven2013} and $\alpha$ is the isotropic dispersivity. In our previous research of conservative tracer cases, we found that the longitudinal dispersion does not play role for determining mixing width because mixing mostly occurs in transverse direction at the interface. Therefore, we use isotropic dispersivity for simplicity. We use $\alpha=0.06$ m obtained by fitting the results of mixing width in conservative experiments using the same grains. $I$ is the identity matrix, $k(x,y,\gamma) = 0.08$ $\si{mM^{-1}.s^{-1}}$ is the reaction rate constant to obtain similar size of reaction zone as in the experiments. The solute concentration at the inlet boundary of the co-flow (at x = 0) was $(C_A,C_B) = (1,0)$ mM for $y>0$ and $(C_A,C_B) = (0,1)$ mM for $y<0$. The inlet boundary condition of the saddle flow was $(C_A,C_B) = (1,0)$ mM at $y=50$ and $(C_A,C_B) = (0,1)$ mM at $y=-50$ mm. For the outlet boundaries ($x=150$ mm for co-flow and $y=\pm50$ mm for saddle flow), we imposed zero gradient for all the species. We used Euler method as a temporal discretisation scheme, and linear interpolation scheme for interpolating face centred values from cell centred values. We varied the velocity field (co-flow) and compression rate (saddle-flow) in the same range as in the experiments.

\section{Results}

\subsection{Diffusion case}
\begin{figure}
    \centering
    \includegraphics[width=\textwidth,height=\textheight,keepaspectratio]{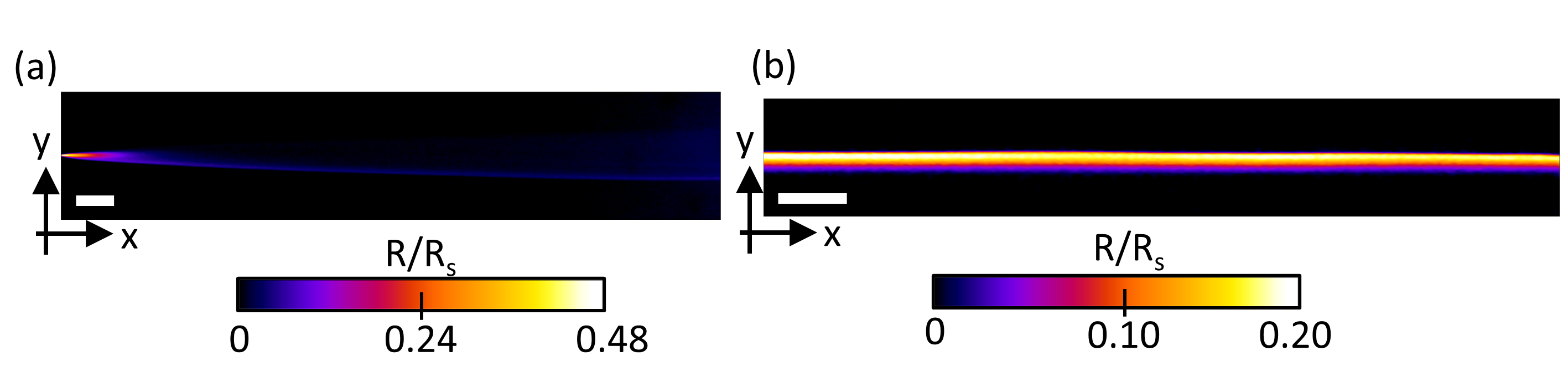}
    \caption{Reaction rate fields at the highest Pe in Hele-Shaw cell. (a) Co-flow. The left edge corresponds to the start of the mixing (upstream). (b) Saddle-flow. The left edge corresponds to the stagnation point. The white bar represents 10 mm.}
    \label{figHImages}
\end{figure}

\begin{figure}
    \centering
    \includegraphics[width=\textwidth,height=\textheight,keepaspectratio]{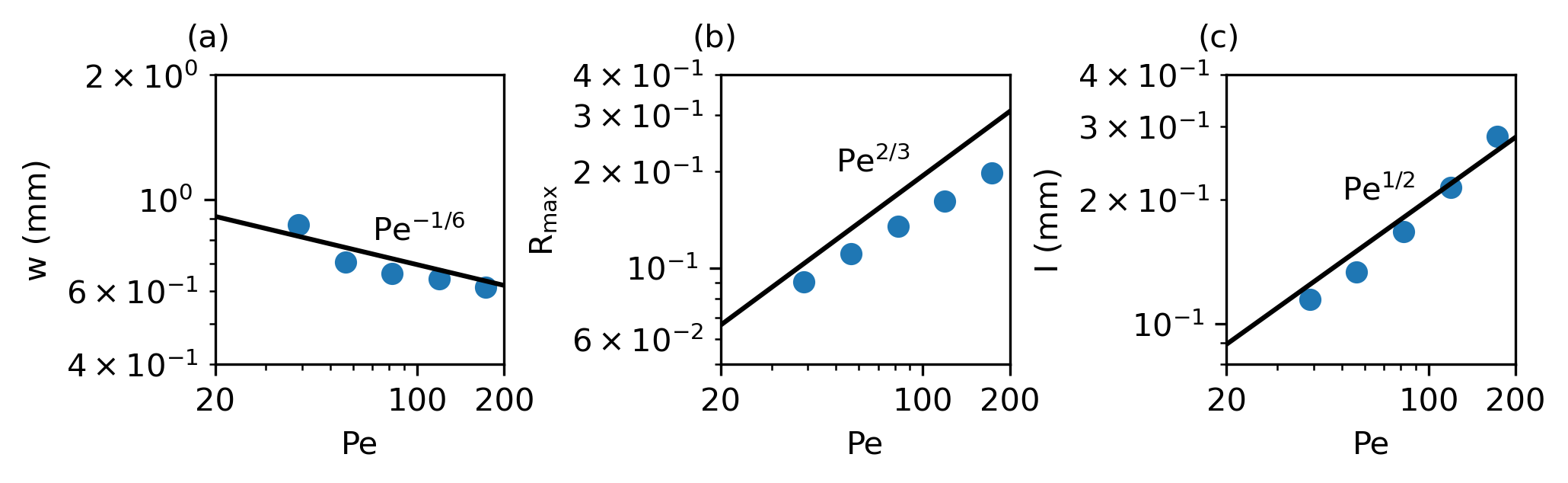}
    \caption{The properties of reaction rate profiles as a function of Pe in saddle flow using the profile averaged over 100 pixels around the stagnation point. The black lines show the scaling laws. (a) Width (b) maximum reaction rate and (c) reaction intensity.}
    \label{figHSaPe}
\end{figure}

In Hele-Shaw cell experiments in co-flow, the reaction rate field showed large variation over the space (Fig.\ref{figHImages}a). From the reaction rate profiles perpendicular to the mixing front (profile along y axis), we quantified the width, maximum reaction rate and reaction intensity as follows. For the width, the standard deviation was calculated by 
\begin{equation}
    s_R=\sqrt{\int x^2P(x)dx - \left(\int xP(x)dx\right)^2},
\end{equation}
where $x$ is the position and $P(x)=R(x)/\int R(x)dx$ with $R(x)$ the normalized reaction rate at the position $x$. The maximum reaction rate of the profile was chosen as $R_{max}$, and the reaction rate was integrated over the measured line, i.e. $\sum(R(x)s_p)$, where $s_p$ is the size of the pixel, to estimate the integral of reaction rate $I_R$. Furthermore, we consider a fluid element that travels from x = 0. The travel time ($\Delta t$) and the position ($x$) are related by $\Delta t = x/v_x$, where $v_x$ is the uniform velocity along x-axis. The travel time is equal to the duration of the reaction within the fluid element, which allows predicting the scaling laws over $\Delta t$ using the theory from diffusion-reaction cases in the previous studies \citep{Larralde1992,Taitelbaum1991}.
As the image shows (Fig.\ref{figHImages}a), the reaction intensity was already large at x = 0 instead of zero because of the difficulty to set the upmost stream boundary. Since we expect the scaling law $I \propto \Delta t^{1/2}$ for early time and $I \propto \Delta t^{-1/2}$ for later time \citep{Larralde1992,Taitelbaum1991}, we compensate this non-zero upstream reaction rate by calculating change in the intensity as $\Delta I = (I^{2}-I^{2}_{x=0})^{1/2}/2$, where $I_{x=0}$ is the intensity at x = 0. After taking the maximum, we calculated the change in intensity  as $\Delta I_{late} = (I^{-2}-I^{-2}_{max})^{-1/2}/2$, where $I_{max}$ is the maximum intensity. For the plotting of $I_{late}$, we calculated the elapsed time after the intensity takes maximum value as $\Delta t_{late} = \Delta t - \Delta t_{max}$, where $\Delta t_{max}$ is the time when the reaction intensity takes maximum value.\\
The quantified reaction properties mostly followed the theoretical scaling laws as expected both in the early time regime, where the diffusion dominates reaction, and in the later time regime, where the reaction dominates diffusion \citep{Larralde1992,Taitelbaum1991}. The exception is that the width showed only the early time regime scaling $w \propto \Delta t^{1/2}$. Because the intensity scaled as $\Delta I \propto \Delta t^{-1/2}$ and the maximum reaction rate as $R_{max} \propto \Delta t^{-2/3}$ at later time, we expected the width to scale as $w \propto \Delta t^{1/6}$ (we expect $I \propto s \times R_{max}$) instead of $w \propto \Delta t^{1/2}$. This discrepancy was probably because the width is sensitive to the tailing part of the reaction rate profile, which has weak image intensity and potentially affected by the long-lasting reaction of the luminol even after the image processing.\\
In saddle-flow, the reaction rate profile was mostly invariant along x-axis (Fig.\ref{figHImages}b). We averaged the reaction rate profile around the stagnation point (100 pixels) to remove noise and quantified the reaction properties in the same way as in the Hele-Shaw cell as a function of Pe. The results followed the expected scaling laws for the case of fast reaction compared to the diffusion (Fig.\ref{figHSaPe}). This indicates that the effect of the tailing of the reaction profile can be ignored in the presence of compression, and also the reaction rate is fast compared to compression rate in this Pe range.

\begin{figure}
    \centering
    \includegraphics[width=\textwidth,height=\textheight,keepaspectratio]{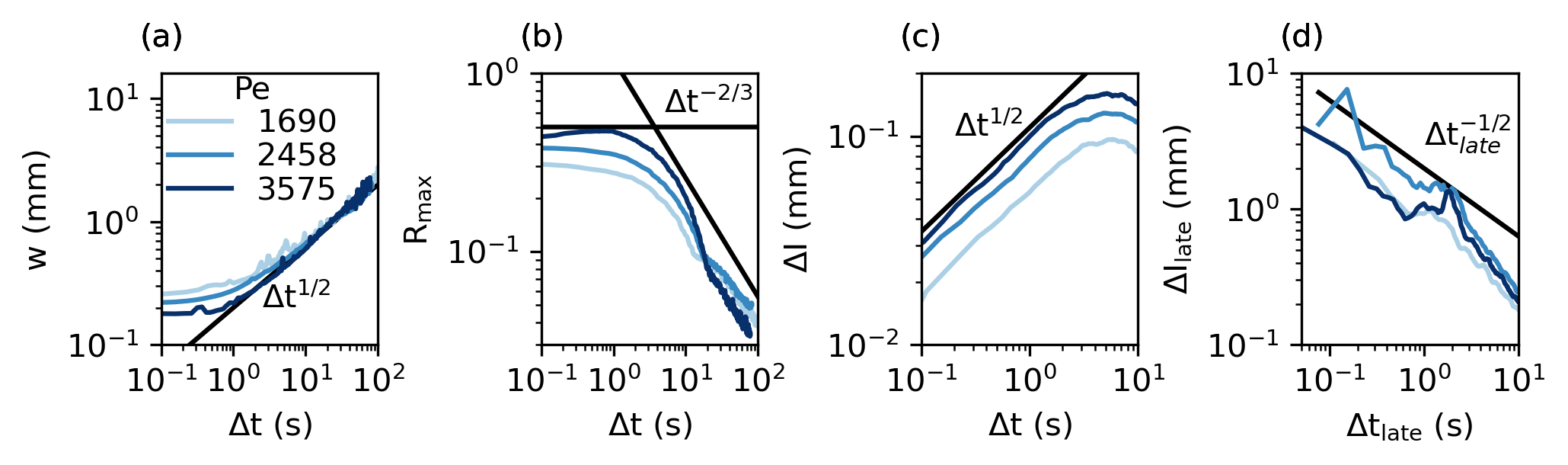}
    \caption{The properties of the reaction rate profiles over the lagrangian time frame in co-flow. The black line show the scaling laws. (a) Width (b) maximum reaction rate (c) change in the intensity from t = 0 and (d) change in the intensity from the maximum intensity. Note that, only for $I_{late}$ plot, $\Delta t$ was calculated from the time when the intensity took maximum value.}
    \label{figHCoTime}
\end{figure}

\subsection{Dispersion case}
\begin{figure}
    \centering
    \includegraphics[width=\textwidth,height=\textheight,keepaspectratio]{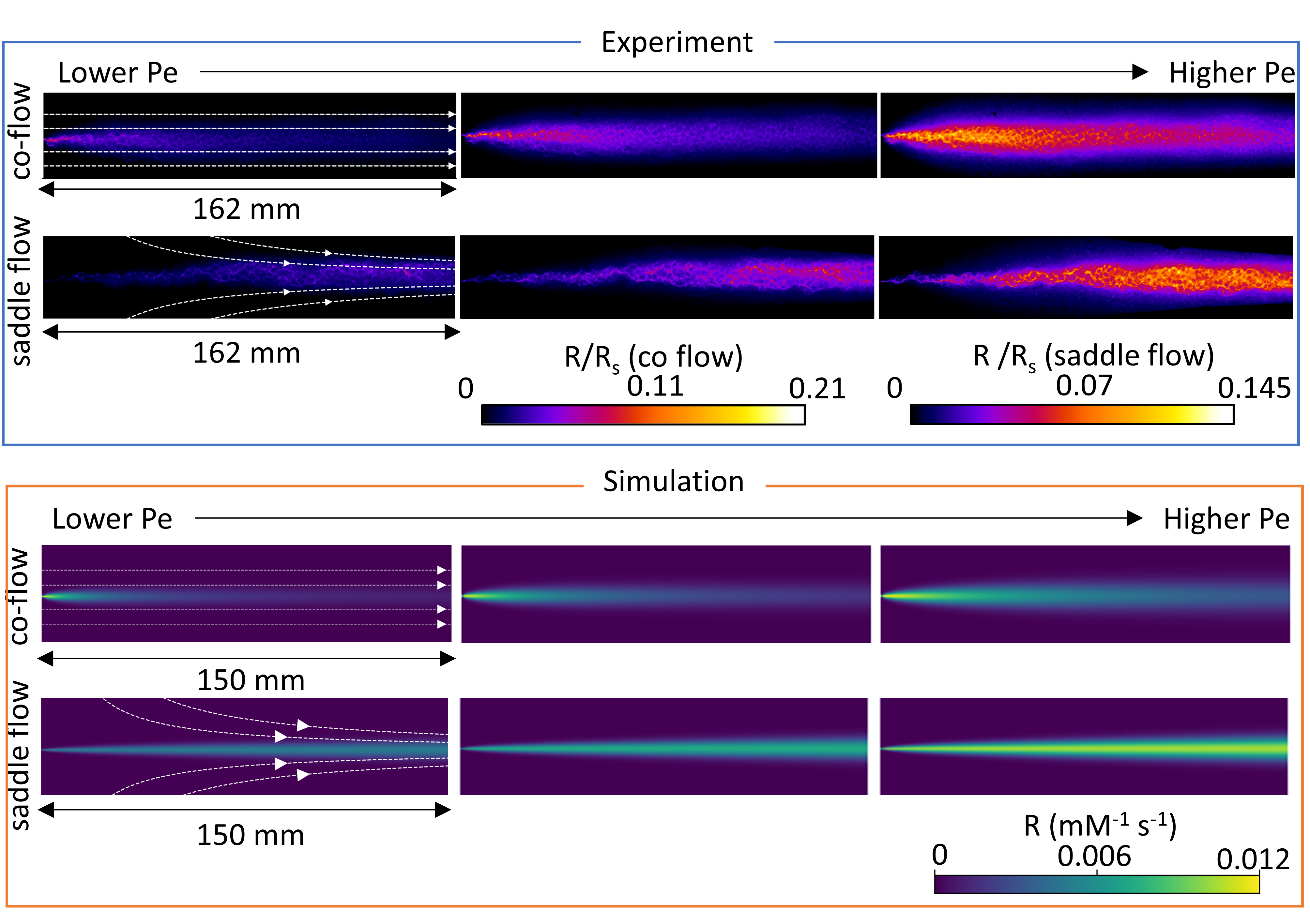}
    \caption{Reaction rate fields at Pe = 799, 1690 and 3575 for co-flow and Pe = 38.8, 82.1 and 173 for saddle-flow (lowest at the left column, highest at the right column) from experiments (top two rows) and from simulations with constant dispersivity and reaction rate constant (bottom two rows). For co-flow, the left edge corresponds to the start of mixing and for saddle-flow, the left edge corresponds to the stagnation point. The white dotted lines in the lowest Pe in experimental images show the streamlines.}
    \label{figPImages}
\end{figure}

\begin{figure}
    \centering
    \includegraphics[width=\textwidth,height=\textheight,keepaspectratio]{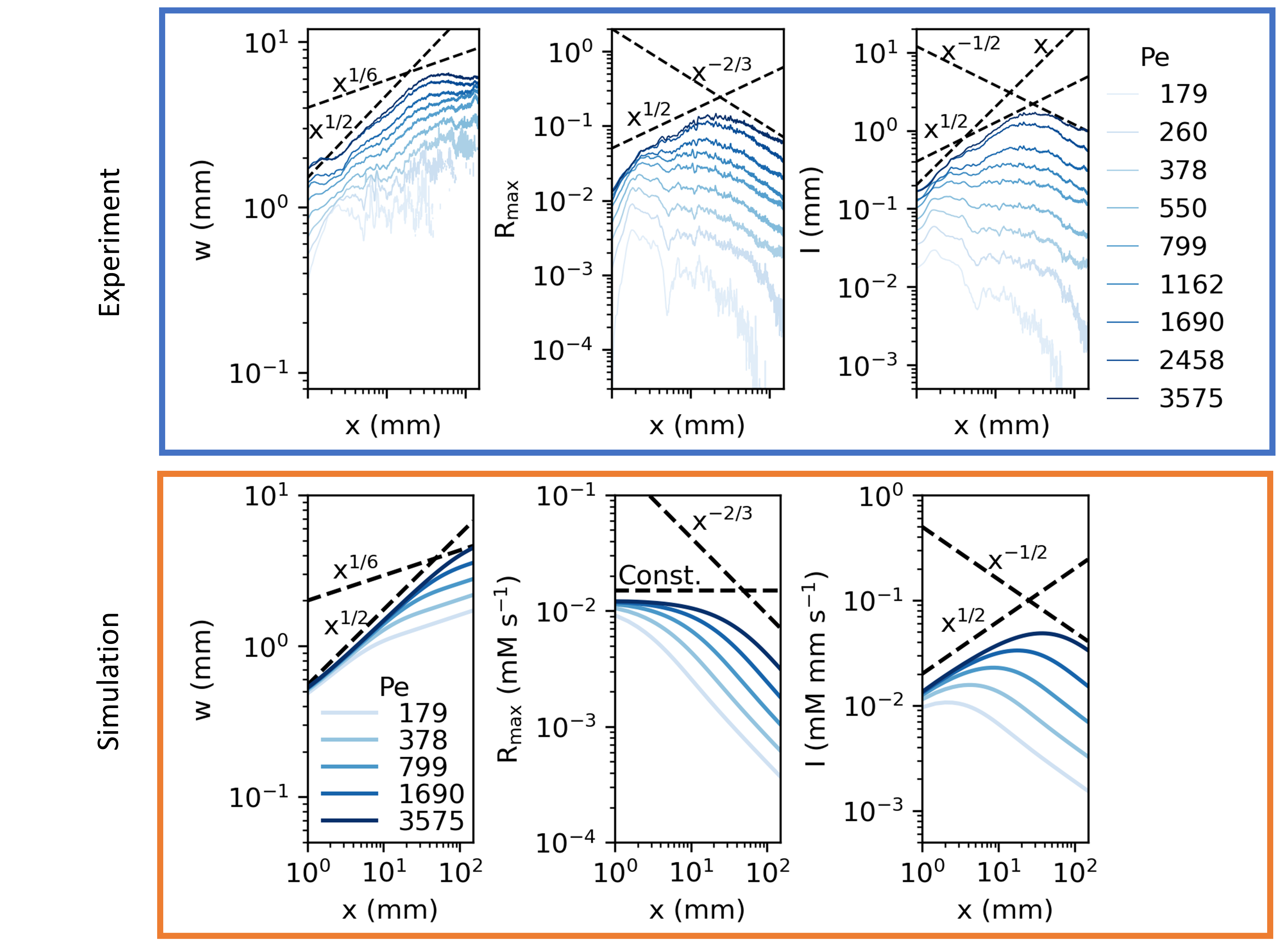}
    \caption{The width (left column), maximum reaction rate (middle column) and reaction intensity (right column) of the reaction rate profiles in co-flow porous media over distance in experimental results (top row) and simulation results (bottom row). For the scaling of Pe, the size was taken around the deflection point at the highest Pe (x = 48 mm for experiment, not detectable at the lowest Pe and x = 16 mm for simulation). We took the maximum of maximum reaction rate and intensity.}
    \label{figCoFlow_dist}
\end{figure}

\begin{figure}
    \centering
    \includegraphics[width=\textwidth,height=\textheight,keepaspectratio]{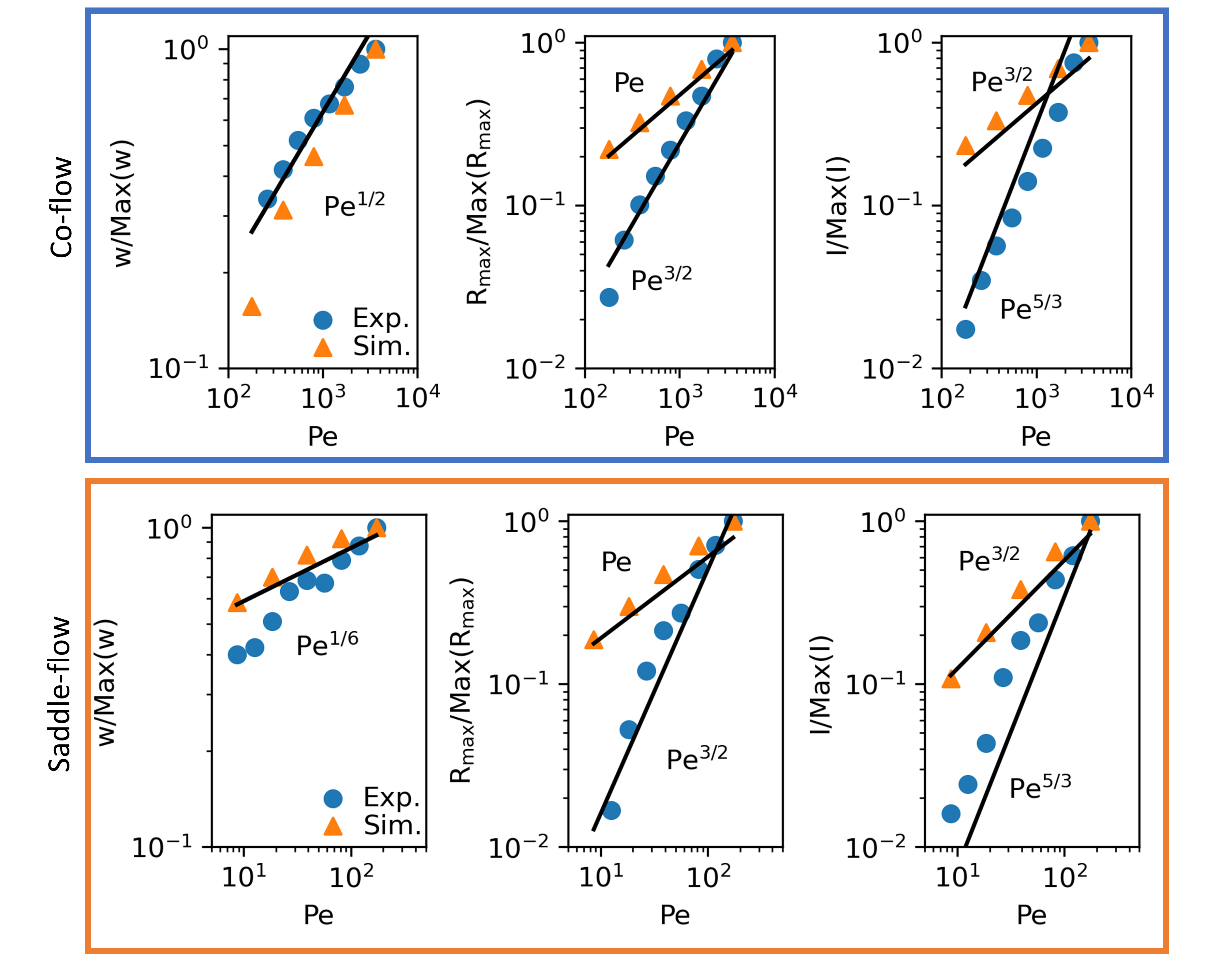}
    \caption{The width (left column), maximum reaction rate (middle column) and reaction intensity (right column) of the reaction rate profiles in co-flow (top row) and saddle-flow (bottom row) porous media over Pe in experiment and simulations. Each properties are normalized by the maximum values both in experiments and simulations (the maximum values are 1 for all the properties in both experiments and simulations). For the co-flow, the width was taken around the deflection point over distance at the highest Pe (x = 48 mm for experiment, not detectable at the lowest Pe and x = 16 mm for simulation). The maximum of maximum reaction rate and the maximum reaction intensity over distance were taken for co-flow. For saddle-flow, the width, maximum reaction rate and intensity were averaged between x = 30 and 35 mm, and those at x = 35 mm were taken in simulations.}
    \label{figPe}
\end{figure}

\begin{figure}
    \centering
    \includegraphics[width=\textwidth,height=\textheight,keepaspectratio]{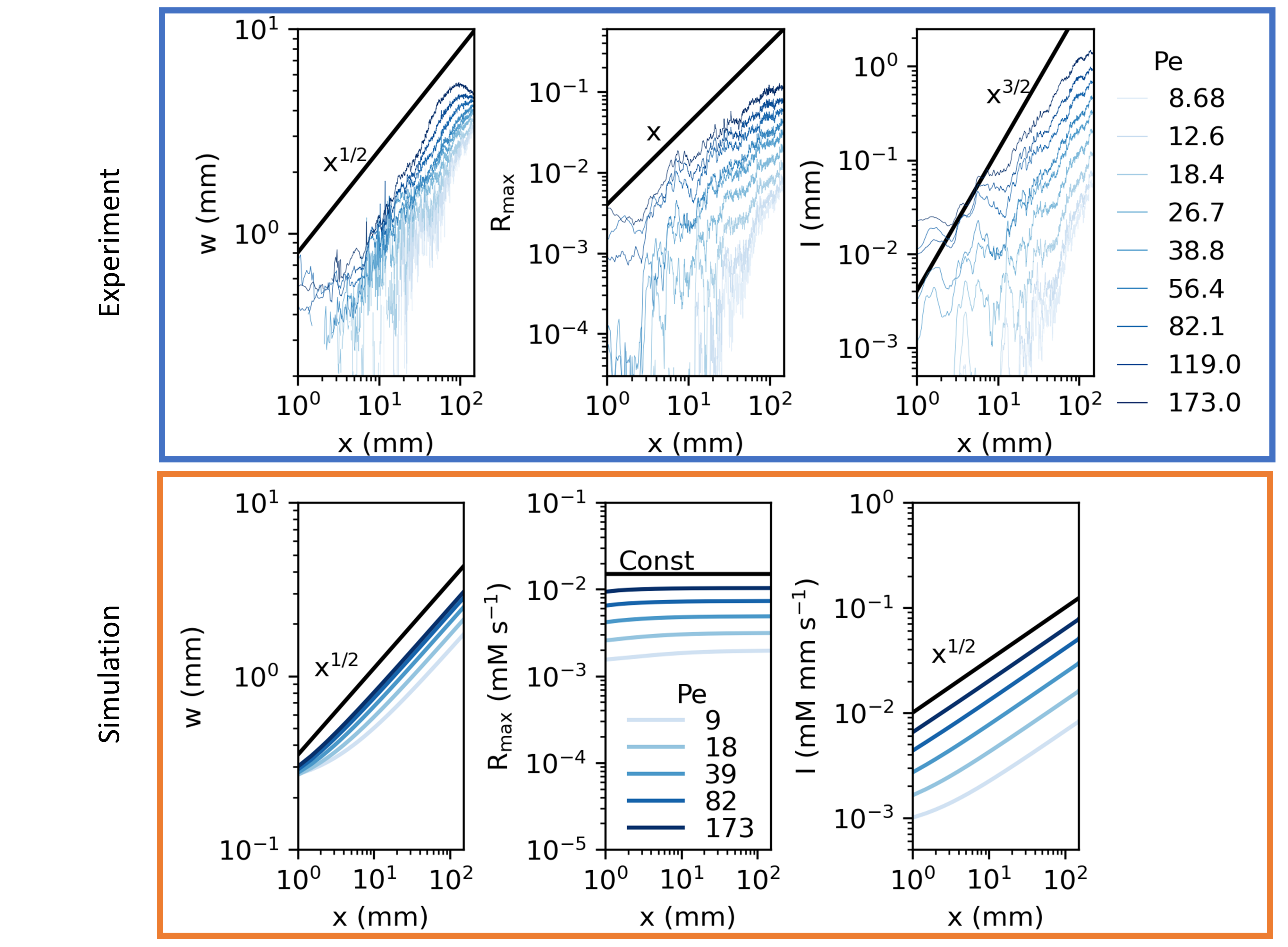}
    \caption{The width (left column), maximum reaction rate (middle column) and reaction intensity (right column) in saddle-flow porous media over distance in experiments (top row) and simulations (bottom row).}
    \label{figSaddleFlow_dist}
\end{figure}

The images successfully captured the light from the luminol chemiluminescence reaction at the interface of the two injected solutions, which were processed to obtain reaction rate field as explained in the method (Fig.\ref{figPImages}). We took the reaction rate profiles perpendicular to the mixing interface, and quantified its width, maximum reaction rate and the integrated reaction rate over the profiles in the same way as the Hele-Shaw cell cases. We averaged the width, maximum reaction rate and intensity over triplicated experiments.\\
In co-flow, under the assumption of fully mixed condition, we expect the same scaling laws as the Hele-Shaw cell by replacing the time and space because the dispersion coefficient is constant under fixed Pe. The numerical simulation supported this prediction (Fig.\ref{figCoFlow_dist} bottom). In the experiment, the maximum reaction rate and reaction intensity before taking the maximum values increased more than in the simulations in most cases (Fig.\ref{figCoFlow_dist} top). In addition, the middle Pe experiments (Pe = 550, 799, 1162) showed almost constant reaction intensity between x = 2 and 50 mm. Such a constant reaction intensity was absent in the numerical simulations. The larger scaling exponent at small x and the constant reaction intensity at middle Pe indicate that pore scale incomplete mixing plays important role in these zones. We also plotted the reaction properties (width, maximum reaction rate and intensity) over Pe by taking average width around the deflection point of the highest Pe (x = 48 mm), the maximum of maximum reaction rate and maximum intensity (Fig.\ref{figPe}). The scaling exponents were again larger compared to the theoretical and numerical expectations assuming the complete mixing, indicating the presence of pore scale process that influences reaction rate fields.\\
In the saddle-flow porous media case, all the properties (width, maximum reaction rate and intensity) kept increasing over the space (Fig.\ref{figSaddleFlow_dist} top) without transition of the scaling laws. This is in contrast to the co-flow case (Fig.\ref{figCoFlow_dist}), where the transition of the scaling laws were observed for all the reaction properties. Thus, such a presence and absence of the transition represents the key role of the fluid acceleration. The maximum reaction rate and the reaction showed larger scaling exponent than those in the numerical simulations over the space (Fig.\ref{figSaddleFlow_dist}). To quantify the Pe dependency, we took the average of the reaction properties between 30 and 35 mm. The scaling exponents of maximum reaction rate and reaction intensity were larger than those in numerical simulations (Fig.\ref{figPe}). These differences between the experiments and numerical simulations indicate that the incomplete pore scale mixing plays an important role.\\

\subsection{Comparison of conservative and reactive experiments}
\begin{figure}
    \centering
    \includegraphics[width=\textwidth,height=\textheight,keepaspectratio]{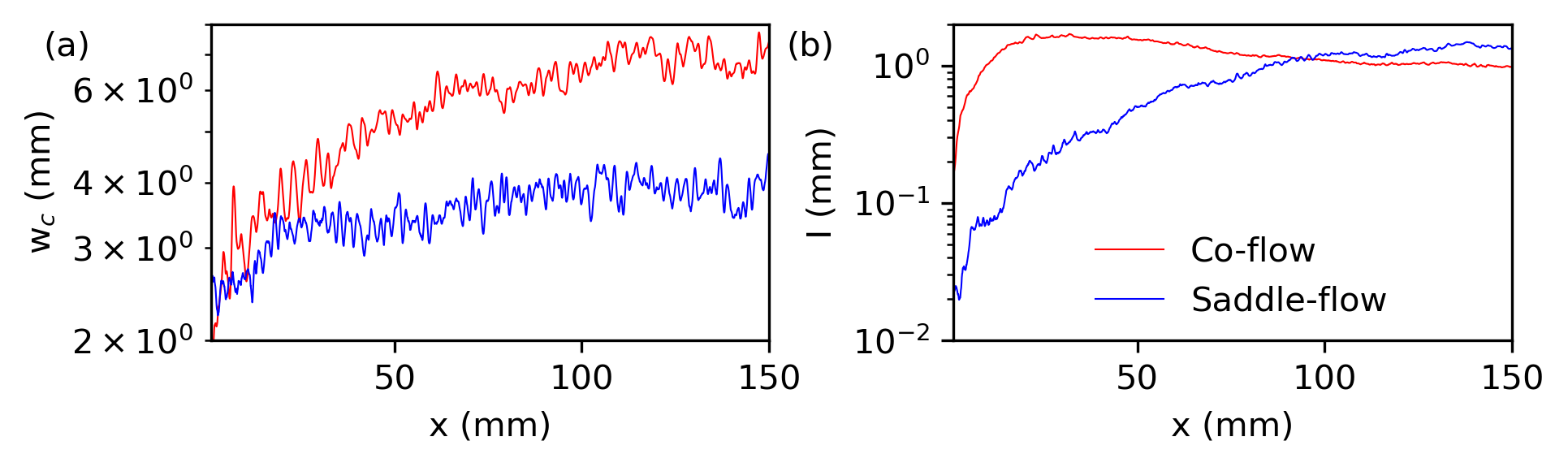}
    \caption{(a) Width of the mixing zone of conservative tracer over distance in co-flow and saddle-flow. The values are the average of all Pe in triplicated experiments. (b) Intensity of reaction at the highest Pe in co-flow and saddle-flow.}
    \label{figConsReact}
\end{figure}

In the experiments using conservative tracer, we quantified the width of the mixing zone by fitting the error function to the concentration profile perpendicular to the mixing interface (more detail in our another paper and in Appendix). Since the width was independent of Pe in conservative tracer experiment (Fig.\ref{figCons} in Appendix), we plot the average width of all the measurement in Fig.\ref{figConsReact}(a). The width of the mixing zone of the conservative tracer kept increasing over distance, from upstream to downstream, both in co-flow and saddle-flow. The width of the mixing zone matched the numerical simulation results, assuming the complete mixing in pore space (details in our another paper). This indicates that even though the conservative case can be modelled by assuming complete mixing, the reactive case in the same porous media should be modelled by including the pore scale incomplete mixing if the reaction is fast.\\
In contrast to the conservative cases, reaction intensity showed different trend (Fig.\ref{figConsReact}b) between co-flow and saddle-flow. In co-flow, the intensity mostly decreased following sharp increased in short distance, while in saddle-flow, the intensity kept increasing. This informs that even though the conservative tracer shows similar trend regarding its mixing zone, the reaction intensity may show opposite trend due to the acceleration of the flow. \\

\section{Discussion}

\subsection{Effect of fluid acceleration}
We first discuss the role of acceleration on reactive transport. In the absence of the acceleration (co-flow), there was the transition of the scaling laws over the space (Fig.\ref{figCoFlow_dist}), whereas there was no transition in the presence of acceleration (saddle-flow) (Fig.\ref{figSaddleFlow_dist}). The previous studies showed that such transition occurs when the diffusion rate becomes slower than the reaction rate in the diffusion-reaction cases \citep{Larralde1992,Taitelbaum1991}. Since our experiments are advection-dispersion-reaction cases, the spreading of the reactants across the mixing interface is controlled by the transverse dispersion. Thus, we compare the characteristic reaction time $\tau_R$ and the characteristic dispersion time $\tau_{disp}$. The characteristic reaction time is obtained $\tau_R = 2/kA_0$ where k is the rate constant, $A_0$ is the bulk reactant concentration. The characteristic time for dispersion $\tau_{disp}$ can be calculated by assuming the larger dispersion compared to the reaction as; $\tau_{disp}=s^2/D_{disp} \propto t \propto x/v$, where $s$ is the size of the mixing zone $s\propto \sqrt{D_{disp}t} = \sqrt{D_{disp}x/v_x}$ by neglecting the reaction. Therefore, $\tau_{disp}$ increases over the distance and the transition occurs when $\tau_{disp}$ becomes small compared to $\tau_R$. On the other hand, for saddle-flow, the characteristic dispersion time is $\tau_{disp}=s^2/D_{disp} \propto \gamma$ where we used $s\propto \sqrt{D_{disp}/\gamma}$. Because $\tau_{disp}$ and $\tau_R$ are constant in saddle-flow, the transition does not occur in saddle flow. This mechanism indicates that the characteristic dispersion time is maintained due to the acceleration, which explains the qualitative difference between with and without acceleration cases.\\ 

\subsection{Enhanced reaction due to folding of lamellae in pore space}
\begin{figure}
    \centering
    \includegraphics{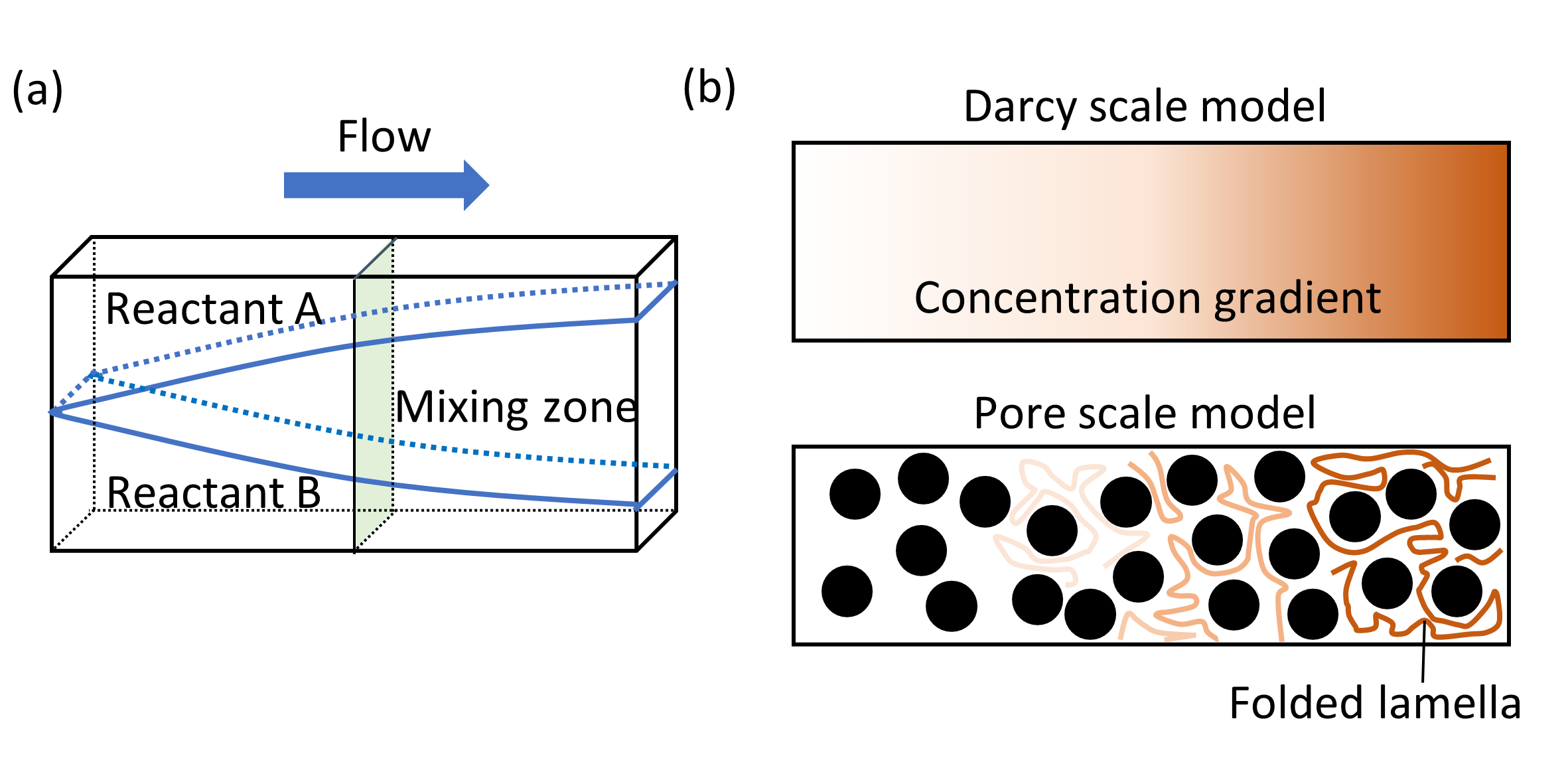}
    \caption{The schematics to show the Darcy scale model and pore scale model. (a) When the reactant A and B flow in parallel in porous media, the mixing and reaction occurs at the interface of A and B. (b) The schematics of the cross section shown as green in (a). In Darcy scale model (top), the concentration of reactant is modelled by continuum scale approach. Thus, the concentration monotonically changes from one side to the other side. In the pore scale model (bottom), the folded lamellae between the grains creates local concentration gradient and local mixing interfaces within the pore space. The concentration inside the lamellae may change over the entire mixing zone.}
    \label{figPoreDarcy}
\end{figure}

\begin{figure}
    \centering
    \includegraphics[width=\textwidth,height=\textheight,keepaspectratio]{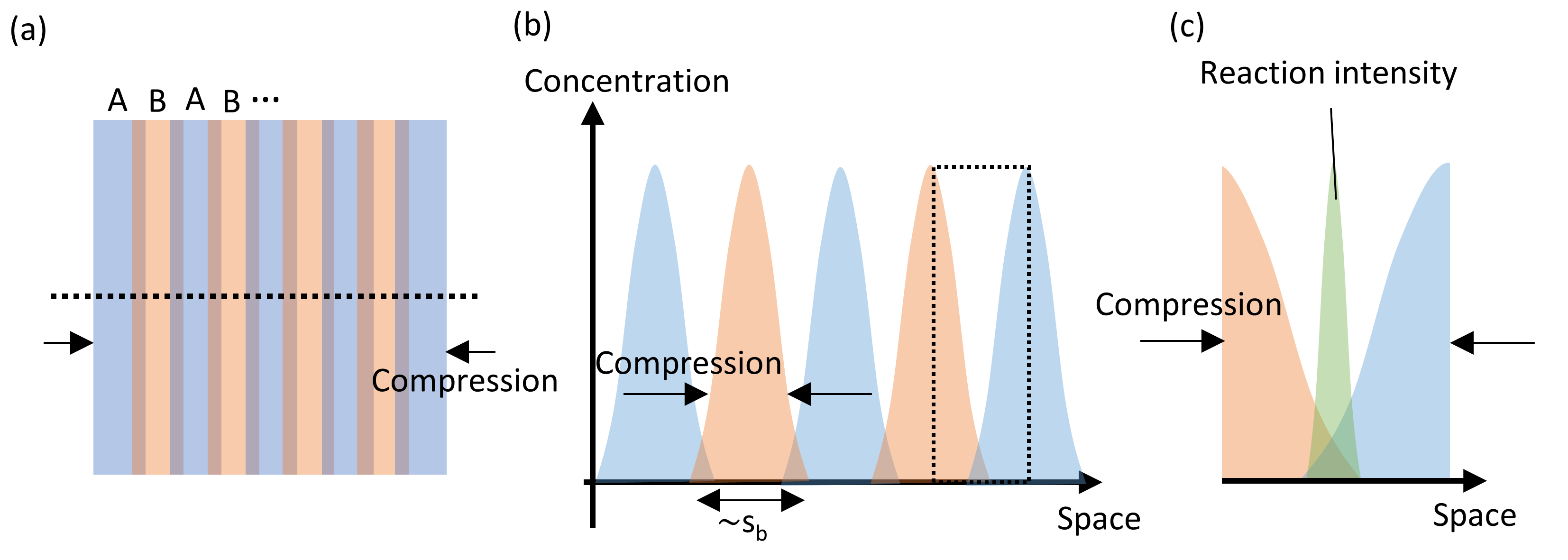}
    \caption{The schematics to show the pore scale mechanisms that enhances reaction. (a) The folded lamellae of species A and B in pore space due to chaotic advection. The compression acts against the lamellae. (b) The concentration profile inside the dotted square in (a), showing the size of the lamellae $s_b$ is determined by the pore scale compression. (c) The closed up of the dotted square in (b), showing that the reaction occurs at the overlapping zone of species A and B. The compression controls the concentration gradients of A and B, which determines the reaction intensity.}
    \label{figSchematic}
\end{figure}

\begin{figure}
    \centering
\includegraphics[width=\textwidth,height=\textheight,keepaspectratio]{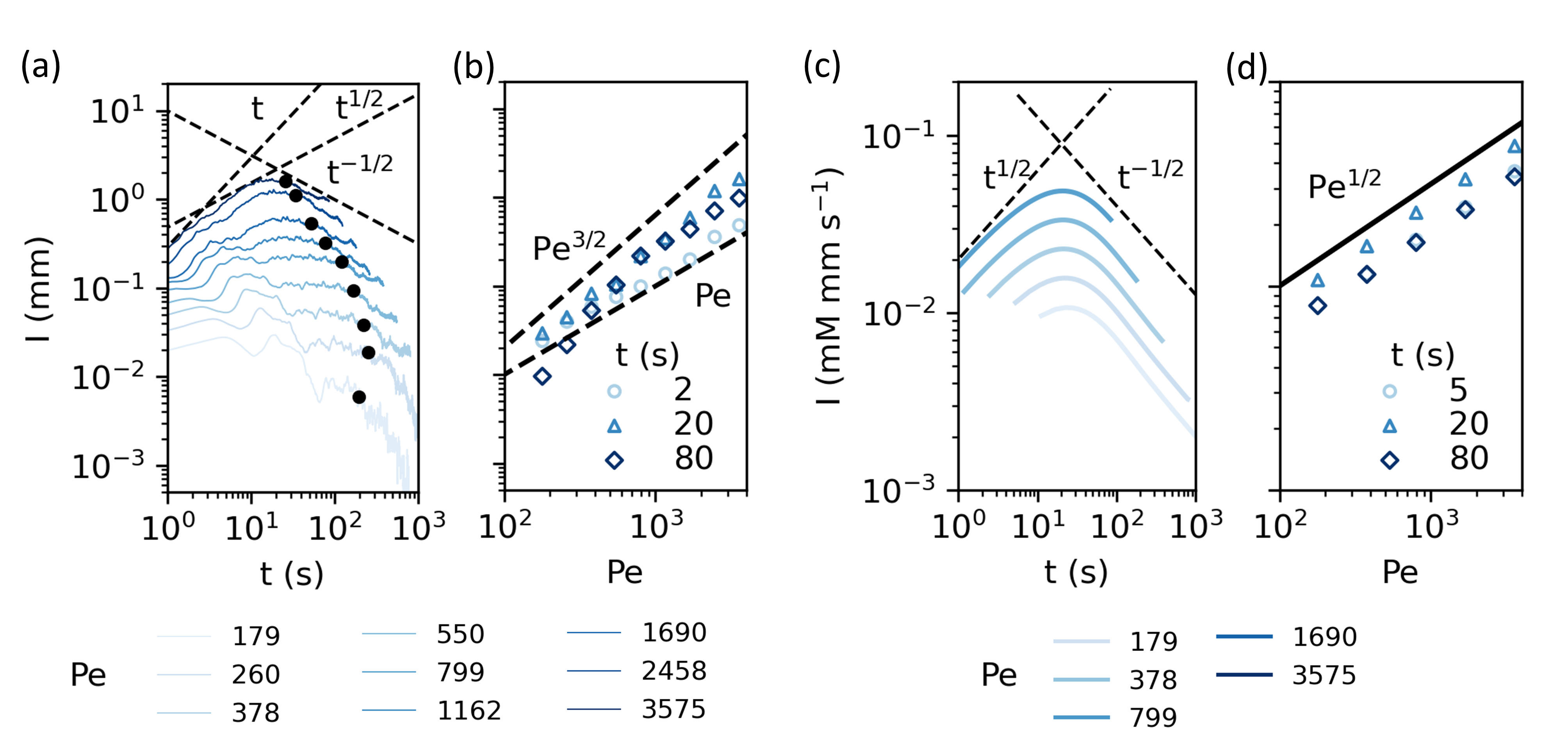}
    \caption{The reaction intensity in co-flow experiments and simulations, over time $t=x/v_x$ where $v_x$ is the Darcy scale velocity, and over Pe. The black dots correspond to the mixing times in each Pe. (a) The reaction intensity over time in the experiments. (b) The reaction intensity over Pe in the experiments at fixed times. (c) The reaction intensity over time in the simulations. (b) The reaction intensity over Pe in the simulations at fixed times.}
    \label{figCoFlowTimePe}
\end{figure}

\begin{figure}
    \centering
\includegraphics{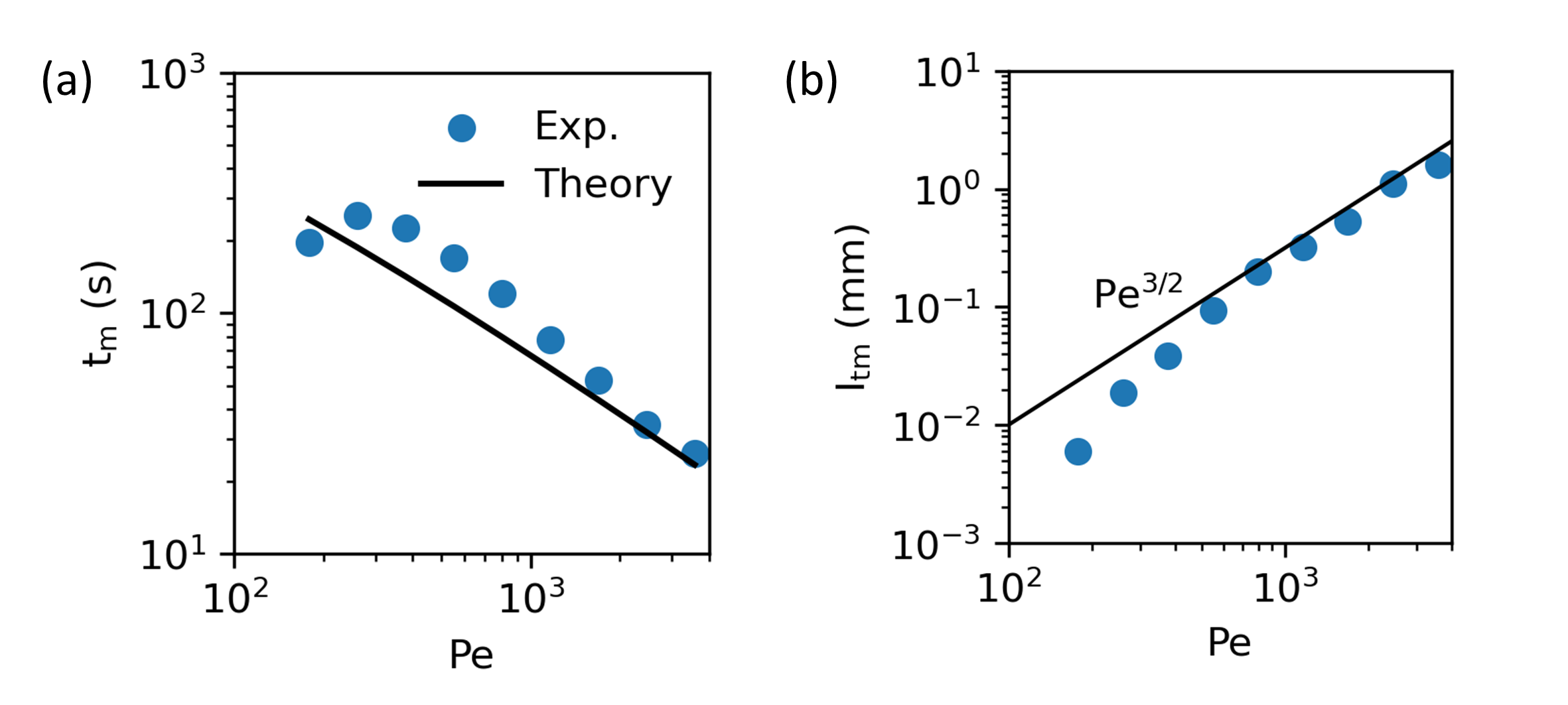}
    \caption{(a) The mixing time of the co-flow porous media experiments over Pe. (b) The reaction intensity at the mixing times over Pe.}
    \label{figMixingTime}
\end{figure}

In the result sections we showed that the reactive transport experimental results could not be reproduced by the numerical simulations. Here we discuss how the pore scale incomplete mixing rectifies the reactive transport. We focus on the reaction intensity because the reaction intensity is the integrated measurement and it is least affected by the local flow conditions compared to the width and maximum reaction rate. Table \ref{tableScalingsMain} summarizes the comparison of scaling exponents between the experiments and simulations. Regarding Pe dependency in co-flow and saddle-flow experiments, scaling exponent are +1 in the experiment compared to the simulations. This indicates that some pore scale mechanism leads to stronger dependency of reaction intensity on velocity. The velocity dependency of reaction intensity is also present in saddle-flow over x because the fluid is accelerating as $v_x=\gamma x$. The experimental scaling exponent was +1 compared to the simulation in saddle-flow over x, which is consistent with the Pe dependency. The pore scale mechanism that leads to +1 exponent for velocity dependency can be explained as follows.\\
In the pore space, the reactants A and B are segregated under the incomplete mixing scenario (Fig.\ref{figPoreDarcy}). The chaotic mixing induces the folding of A and B, which leads to layers of A and B in pore space (Fig.\ref{figSchematic}a). When velocity increases, the compression rate due to chaotic mixing increases proportionally. The increase of compression rate results in decrease in the thickness of each layer of A and B (Fig.\ref{figSchematic}b). Since the layer thickness is determined by the balance of diffusion and compression, the thickness scales similar as the Batchelor scale as $l \propto \mathrm{Pe}^{-1/2}$, where $l$ is the layer thickness. Thus, the number of interfaces between the layers $N$ scales as $N \propto 1/l \propto \mathrm{Pe}^{1/2}$. The increase in compression also enhances the concentration gradient of A and B (Fig.\ref{figSchematic}c). This leads to the enhancement of reaction at the interfaces as $I \propto \mathrm{Pe}^{1/2}$ when the reaction is faster than the diffusion in the same way as the enhancement of reaction at stagnation points. Such an increase in the number of interface and the enhancement of the reaction at the interfaces drive the pore scale reaction as $I N \propto \mathrm{Pe}$. \\
In addition to these pore scale mechanisms, mixing occurs due to dispersion and Darcy scale compression, that are simulated by the numerical simulations. The mixing in Darcy scale controls the concentration profiles of reactants and entire mixing zone in Darcy scale. The concentration profile in Darcy scale determines the concentration of each folded layer A and B in pore scale, and the mixing zone in Darcy scale rectifies the zone where the pore scale mixing occurs. Therefore, we may multiply the scaling law of Darcy scale reaction intensity and that of pore scale enhanced reaction intensity. This leads to exponent +1 by $I N \propto Pe$ due to pore scale mechanisms compared to Darcy scale simulation.\\
The effects of incomplete mixing on reactive transport was also observed without variation of velocity. In the co-flow experiment, the Darcy scale velocity is constant over x, but the results showed different trends over x compared to the simulations (Table \ref{tableScalingsMain}, Fig.\ref{figCoFlow_dist}). The large increase of the reaction intensity, where the scaling exponent is close to 1, was observed only at the first few grains  (Fig.\ref{figCoFlow_dist}). This can be attributed to the exponential elongation of the mixing interface due to chaotic mixing \citep{Lester2013,Lester2014,Lester2016}. We also observed the almost constant reaction intensity in wide range of x in middle Pe. To understand the underlying mechanisms, we plot the reaction intensity over time by the relation $t=x/v_x$ (Fig.\ref{figCoFlowTimePe}a,c). The maximum reaction intensities are close to 20 s in all Pe, which is consistent with the numerical simulations. This suggests that $\tau_{disp}\propto t$ and $\tau_{R}$ are not affected by pore scale incomplete mixing when the reaction intensity takes the maximum value because these two characteristic times control when the reaction intensity takes maximum value as discussed in the previous section. This is consistent with the saddle-flow experimental results because the absence of the transition of the scaling laws in saddle-flow could be also interpreted without inferring pore scale processes.\\
For the interpretation of the constant reaction intensity at middle Pe over time, we hypothesize that the scaling of $t^{-1/2}$ resulting from Darcy scale at later time regime, as observed in the numerical simulations (Fig.\ref{figCoFlowTimePe}c), balances the increase in the reaction rate due to pore scale compression (Fig.\ref{figSchematic}). To check this hypothesis, we theoretically calculate the time when the pore scale reaction rate saturates by reaching the maximum number of interfaces. This time would correspond to the mixing time, when the size of lamellae reaches equilibrium under constant compression rate \citep{Villermaux2012}:
\begin{equation}
    t_m = \frac{1}{2\gamma_p}\text{ln}\left(\frac{\gamma_p s_0^2}{D}\right)
    \label{eqMixingTime}
\end{equation}
where $t_m$ is mixing time, $s_0$ is the initial lamellae size, $D$ is diffusion coefficient and $\gamma_p$ is pore scale compression rate. $\gamma_p$ can be calculated by the pore scale velocity $v_p$, grain diameter $d$ and lyapunov exponent $\lambda$ as $\gamma_p = \lambda v_p/d$. We calculated the mixing time $t_m$ by setting $d$ as 2 mm and $\lambda$ as 0.15, which is between the case of random pore network (0.12) and random packing of spherical beads (0.21) \citep{Heyman2020}. We assumed 2 mm for the initial lamellae size (same as the size of the grain). From the experiments, we manually picked up the mixing times when the scaling of $t^{-1/2}$ starts (Fig.\ref{figCoFlowTimePe}a). The comparison of experimental mixing times and theoretical mixing times showed good agreement (Fig.\ref{figMixingTime}a). This suggests that the transition of the scaling law from the exponent 0 to -1/2 in co-flow experiments corresponds to the time when the pore scale interface reached its maximum length. The Pe dependency of reaction intensity at the mixing time $I_{tm} \propto \text{Pe}^{3/2}$ was the same as that of the maximum reaction intensity (Fig.\ref{figMixingTime}b). This can be again interpreted by the pore scale compression of lamellae (Fig.\ref{figSchematic}). Therefore, the constant reaction intensity before mixing time (scaling exponent of 0 in Table.\ref{tableScalingsMain}, Fig.\ref{figCoFlowTimePe}) was achieved by the balance between the pore scale enhancement of reaction (Fig.\ref{figSchematic}) and Darcy scale reduction of reaction rate.\\
We further checked the Pe dependency of reaction intensity at different times (Fig.\ref{figCoFlowTimePe}b,d). The dependency on Pe was weaker in early time ($\mathrm{Pe}$ at t = 2 s) than later times ($\mathrm{Pe}^{3/2}$ at t = 20 s and 80 s) (Fig.\ref{figCoFlowTimePe}b,d). In contrast, it was always $\mathrm{Pe}^{1/2}$ in the simulations. This is again consistent with the above discussion in that the pore scale incomplete mixing were still in progress in early times, and thus the scaling exponent was closer to the simulation in the early times. The results also showed that the Pe dependency changed between 20 s and 80 s only at lower Pe in the experiments. This is because the mixing time was larger in low Pe, and thus the pore scale process still had influences on the Pe dependency in low Pe.

\begin{table*}
\centering
\caption{\label{tableScalingsMain}Exponents of scaling laws of reaction intensity. The $x_{small}$ and $x_{large}$ corresponds to the scaling before and after taking the maximum reaction intensity, respectively. In the co-flow experiment in $x_{small}$, the scaling exponent was between 0 and 1 and in $x_{large}$, the exponent was between -1/2 and 0.}
%\begin{ruledtabular}
\small
\setlength\tabcolsep{2pt}
\begin{tabular}{cccccc}
\hline
 &\multicolumn{3}{c}{Co-flow}&\multicolumn{2}{c}{Saddle-flow}\\
 \cmidrule(lr){2-4}\cmidrule(lr){5-6}
&$x_{small}$&$x_{large}$&$Pe$&$x$&$Pe$\\ \hline
 Simulation&$1/2$&$-1/2$&$1/2$ &$1/2$&$2/3$ \\
 Experiment &[0,1]&$[-1/2,0]$&$3/2$&$3/2$&$5/3$\\
 \hline
\end{tabular}
%\end{ruledtabular}
\label{tableScalingsMain}%
\end{table*}

\section{Conclusion}
We experimentally and numerically investigated how incomplete mixing impacts the reactive transport with and without acceleration of flow in Darcy scale. The experiments utilized the packed bed of fluorinated ethylene propylene (FEP), which has the refractive index very close to that of water. By inducing the luminol chemiluminescence reaction in the packed bed, we visualized the reaction rate field of bimolecular second order $A+B\to C + \text{photon}$ reaction. We found that the reaction rate keeps increasing from upstream to downstream under the acceleration, whereas the reaction rate first increases and then decreases without acceleration, even though the conservative tracer experiments showed increase in the mixing interface in both cases. This is because the acceleration sustained the characteristic time of dispersion and reaction at the mixing interface, while the characteristic time of dispersion increases without acceleration. The comparison to the numerical simulations showed that pore scale incomplete mixing enhanced the dependency of reaction rate on flow velocity. To explain this result, we suggested the mechanistic models based on chaotic mixing in pore space. One is that the compression controls the size of the folded layers of reactants, and the other is that the concentration gradient at the edge of the layers are determined by the compression. These two mechanisms successfully explained the observed dependency of reaction rate on velocity. Furthermore, we found that the pore scale incomplete mixing leads to constant reaction rate over distance without variation of velocity. The results was explained by the balance between the Darcy scale reduction of reaction rate and pore scale enhancement of reaction due to folding of reactants and the development of mixing interface due to chaotic mixing.\\
In contrast to the previous studies that observed the reaction product in reactive transport experiments in uniform flow, our experimental results of reaction rate field under accelerating flow is unique and, in combination with simulations, brought new mechanistic understanding on underlying mechanisms of how incomplete mixing impact Darcy scale reaction through the chaotic mixing. More detailed observation of pore scale reacting process would be necessary to validate the proposed pore scale mechanisms.

\begin{acknowledgments}

\end{acknowledgments}

\appendix
\section{Lamella mixing theory for bimolecular second-order reaction}
In this section, we derive the lamellar mixing theory of mixing to derive approximate solutions for coupled mixing and reaction at a stagnation point. When approaching the stagnation point, fluid elements deform
due to elongation and compression in the flow field. The compression rate $\gamma$ is defined as:
\begin{equation}
\label{eqGamma}
    \gamma=-\frac{1}{\tilde{\delta}}\frac{d{\tilde{\delta}}}{dt},
\end{equation}
with $\widetilde{\delta}$ is the width of a fluid element in the direction of compression. Stagnation points are characterized by a locally constant compression rate $\gamma$, such that the width $\widetilde{\delta}$ decays exponentially over time.
\begin{equation}
\label{eqExponential}
    \widetilde{\delta}=\widetilde{\delta}_0 e^{-\gamma t},
\end{equation}
where $\widetilde{\delta}_0$ is the initial lamella size. Following \citet{Ranz1979}, we assume that concentration gradients along the direction of elongation are negligible and that concentration distributions are thus mainly driven by mass transfer in the direction of compression. In a Lagrangian framework aligned with the directions of elongation and compression, the concentration of a conservative species $C_A$ is governed by the compression diffusion equation :
\begin{equation}
\label{eqDiff}
    \frac{\partial C_A}{\partial t} - \gamma x \frac{\partial C_A}{\partial x}=D\frac{\partial^2 C_A}{\partial x^2}
\end{equation}
where $x$ is the Lagrangian coordinate along the compression direction. For a bimolecular reaction reaction $A+B \to C$, the above equation becomes:
\begin{equation}
\label{eqReact}
    \frac{\partial C_A}{\partial t} - \gamma x \frac{\partial C_A}{\partial x}=D\frac{\partial^2 C_A}{\partial x^2}-kC_AC_B
\end{equation}
where $C_A$ and $C_B$ are concentrations of reactants, normalized by the initial concentration $C_0$, $k$ is the reaction constant for dimensionless concentrations $k= \tilde{k}C_0$ with the rate constant $\tilde{k}$ in units of inverse of concentration and time. 
This equation can be transformed to a diffusion-reaction equation using the following change of variables \citep{Ranz1979}:
\begin{equation}
\label{eqRanz}
    \theta=\int_{0}^{t}{d\tau D/{\tilde{\delta}(\tau)}^2}, z=x/\tilde{\delta}
\end{equation}
where $\theta$ is called warped time. In these non-dimensional variables, Eq.(\ref{eqReact}) reduces to a diffusion-reaction equation:
\begin{equation}
\label{eqDaADE}
    \frac{\partial C_A}{\partial \theta} =\frac{\partial^2 C_A}{\partial z^2} - \mathrm{Da} C_A C_B\delta^2
\end{equation}
with the Damköhler number defined as $\mathrm{Da}=k{\tilde{\delta}}_0^2/D$ and $\delta = \tilde{\delta}/\tilde{\delta_0}$.
For exponential compression (Eq.\ref{eqExponential}), the warped time is: 
\begin{equation}
\label{eqTheta}
    \theta=\frac{D}{\gamma{\tilde{\delta}}_0^2}\frac{1}{2}\left(e^{2\gamma t}-1\right)
\end{equation}
This equation is analytically tractable in two limiting regimes, where chemistry is relatively fast or relatively slow compared to the mixing time scales. The mixing timescale is represented by the Peclet number defined as $\textrm{Pe}_{\gamma}=\gamma \widetilde{\delta}_0/D$. These regimes are determined by the ratio $\mathrm{Da}/\mathrm{Pe}_{\gamma}=k \gamma^{-1}$, which compares the characteristic reaction time $k^{-1}$ to the compression time $\gamma^{-1}$.

\subsection{Large $\mathrm{Da}/\mathrm{Pe}_{\gamma}$ regime}
When $\mathrm{Da}\gg \mathrm{Pe}_{\gamma}$, reaction time is small compared to the compression time. Interpenetration of reactants is limited since reaction rapidly deplete their concentration in the mixing zone. Therefore, in the region where the concentration of substance B is large, A is small and \textit{vice-versa}. Following \citet{Larralde1992} and  \citet{Bandopadhyay2017}, the concentration of $C_A$ and $C_B$ can be written in terms of a conservative component $F=C_B-C_A$ and a perturbation $g$, such that $C_A=g$ and $C_B=F+g$. $F$ follows the equation: 
\begin{equation}
\label{eqFdiff}
    \frac{\partial F}{\partial \theta}=\frac{\partial^2F}{\partial z^2}
\end{equation}
which leads to the solution:
\begin{equation}
\label{eqF}
    F= \textrm{erf}\left(\frac{z}{\sqrt{4\theta}}\right)
\end{equation}
Inserting $C_A=g$ and $C_B=F+g$ into Eq.(\ref{eqDaADE}), we obtain:
\begin{equation}
\label{eqgOrig}
    \frac{\partial g}{\partial \theta}=\frac{\partial^2g}{\partial z^2}-\mathrm{Da} g\left(\textrm{erf}\left(\frac{z}{\sqrt{4\theta}}\right)+g\right)\delta^2
\end{equation}
The term proportional to $g^2$ can be neglected\cite{Larralde1992} and the error function in the mixing zone can be linearized. This leads to the approximation:
\begin{equation}
\label{eqg}
    \frac{\partial g}{\partial \theta}\approx\frac{\partial^2g}{\partial z^2}-\mathrm{Da} g\frac{z}{\sqrt{\pi \theta}}\delta^2
\end{equation}
For the stationary conditions that develop at the stagnation point, this equation becomes:
\begin{equation}
\label{eqgSt}
    \frac{\partial^2g}{\partial z^2}-\mathrm{Da} g\frac{z}{\sqrt{\pi \theta}}\delta^2=0
\end{equation}
This is an Airy differential equation. We require the solution to have the form $\psi\theta^{\alpha}\textrm{Ai}(\lambda z/\theta^{1/6})$ following \citet{Larralde1992} and \citet{Bandopadhyay2017}, where Ai is the Airy function. By equating the second derivative term of Eq.(\ref{eqgOrig}) and the second term of RHS in Eq.(\ref{eqg}), we find $\psi=\textrm{Da}^{-1}\lambda^2\delta^{-2/3}$ and $\alpha=-1/3$, where $\lambda=\textrm{Da}^{1/3}\pi^{-1/6}$. The reaction rate in nondimensional space $\widetilde{R}(z,\theta)$ is approximated by the second term in Eq.(\ref{eqg}):
\begin{equation}
\label{eqRz}
\widetilde{R}(z,\theta) 
    \approx \pi^{-5/6}\textrm{Da}^{1/3}\theta^{-2/3}\delta^{2/3}\left(\frac{\lambda z \delta^{2/3}}{\theta^{1/6}}\right)\\
    \textrm{Ai}\left(\frac{\lambda z \delta^{2/3}}{\theta^{1/6}}\right)
\end{equation}

The reaction rate in dimensional space is derived by multiplying $dt/d\theta$ and replacing $z$ by $x$ using Eq.(\ref{eqRanz}), and inserting $\delta=e^{-\gamma t}$ and $\theta = De^{2\gamma t}/2\gamma \tilde{\delta_0}^2$ at large times as:\\

\begin{eqnarray}
\label{eqRx}
\widetilde{R}(x,t) &\approx &2^{2/3}\pi^{-5/6}\textrm{Da}^{1/3}\textrm{Pe}^{2/3}\left(2^{1/6}\pi^{-1/6}\textrm{Da}^{1/3}\textrm{Pe}^{1/6}\frac{x}{\widetilde{\delta}_0}\right)\nonumber\\
&& \times\textrm{Ai}\left(2^{1/6}\pi^{-1/6}\textrm{Da}^{1/3}\textrm{Pe}^{1/6}\frac{x}{\widetilde{\delta}_0}\right)
\end{eqnarray}

The above equation has the form of 
\begin{equation}
\label{eqRscale}
    R(x,t) = R_{max}f\left(\frac{x}{\widetilde{\delta}_0w}\right)
\end{equation}
where we define nondimensional width of reaction zone $w = \widetilde{w}/\widetilde{\delta}_0$, nondimensional maximum reaction rate $R_{max} = \widetilde{R}_{max}\widetilde{\delta}_0^2/D$ and nondimentional reaction rate $R(x,t)=\widetilde{R}(x,t)\widetilde{\delta}_0^2/D$. The tilde represents dimensional variables. We thus obtain the scaling forms for nondimentional variables as:

\begin{equation}
\label{eqS}
    w \propto \mathrm{Da}^{-1/3}\mathrm{Pe}_{\gamma}^{-1/6}
\end{equation}
and
\begin{equation}
\label{eqR}
    R_{max} \propto \mathrm{Da}^{1/3}\mathrm{Pe}_{\gamma}^{2/3}
\end{equation}
The integral of the reaction rate over the direction of compression (the reaction intensity $I$) thus scales as: 
\begin{equation}
\label{eqI}
    I \propto s_RR_{max} \propto \mathrm{Pe}_{\gamma}^{1/2}
\end{equation}

\subsection{Small $\mathrm{Da}/\mathrm{Pe}_{\gamma}$ regime}

When $\mathrm{Da}\ll \mathrm{Pe}_{\gamma}$, the reaction time is large compared to the compression time. The interpenetration of the reactants in the mixing zone is important since reaction slowly deplete their concentration. In this regime, the concentration profiles for $C_A$ and $C_B$ are expected to be close to that of conservative species. By solving Eq.\ref{eqDiff}, $C_A$ and $C_B$ are obtained as:
\begin{equation}
\label{eqCACB}
    C_A= \frac{1}{2}\left(1+\text{erf}\left(\frac{x}{\sqrt{2D/\gamma}}\right)\right), 
    C_B= \frac{1}{2}\left(1-\text{erf}\left(\frac{x}{\sqrt{2D/\gamma}}\right)\right)
\end{equation}
The nondimentional reaction rate $R= \textrm{Da} C_A C_B$ is calculated as 
\begin{equation}
\label{eqRlowDa}
    R = \frac{1}{4}\textrm{Da}\left(1-\textrm{erf}\left(\frac{x}{\sqrt{2D/\gamma}}\right)^2\right)
\end{equation}
Hence, the nondimentional reaction width scales as,
\begin{equation}
\label{eqSLowDa}
    w \propto \mathrm{Pe}_{\gamma}^{-1/2}
\end{equation}
the nondimentional maximum reaction rate as:
\begin{equation}
\label{eqRLowDa}
    R_{max} \propto \textrm{Da}
\end{equation}
and the nondimentional reaction intensity as:
\begin{equation}
\label{eqILowDa}
    I \propto \textrm{Da} \mathrm{Pe}_{\gamma}^{-1/2}
\end{equation}

\section{Predictions assuming pore scale complete mixing}
Here we discuss the scaling laws for the properties of the reaction; width of the reacting zone $w$, maximum reaction rate $R_{max}$ and the reaction intensity $I$ as summarized in Table.\ref{tableScalings}. First, we consider the case of Hele-Shaw cell with co-flow, where the reaction species are injected in parallel. We consider the Lagrangian framework by relating space $x$ and time $t$ as $t = x/v$, where $v$ is the uniform fluid velocity. We can use the scaling laws derived in diffusion-reaction system over time \citep{Larralde1992,Taitelbaum1991}. At closer to the inlet, we may apply the scaling laws for the early time regime in diffusion-reaction system. In this regime, the concentration profile of reactant species can be approximated by that of the conservative species because the reaction is slow. This gives $w \propto (Dt)^{1/2} \propto t^{1/2}$, $R_{max}:const$, $I \propto wR_{max} \propto t^{1/2}$. At far from the inlet, the later time regime of the reaction-diffusion system gives $w \propto t^{-1/6}$, $R_{max} \propto t^{-2/3}$, $I \propto t^{-1/2}$. The transition may occur when the diffusion time balances reaction time. The characteristic time of the diffusion is $\tau_D=s^2/D$, where $s$ is the size of the mixing zone to access the concentration gradient of reactants as $\lambda C = C_0/s$, where $C_0$ is the bulk concentration. The characteristic time for the reaction is $\tau_R = 1/kA$. At the transition time $t=t_c$, the size of the mixing zone can be approximated as $s\sim \sqrt{Dt_c}$. When the reaction rate balances diffusion rate as $\tau_R \sim \tau_D$, the transition time can be written as $t_c \sim 1/kA$. \\
For the Hele-Shaw cell in saddle flow, compression rate is constant over the space. By lamellar mixing theory, the scaling laws were derived in our previous study. The scaling laws vary depending on whether the reaction is faster than the compression-enhanced diffusion or not. At the transition, the size of the mixing zone can be approximated by $s_c=\sqrt{D/\gamma}$, where $\gamma$ is the compression rate given by the velocity gradient. The characteristic time of diffusion is $\tau_D=s_c^2/D=1/\gamma$. On the other hand, the characteristic time of reaction is $\tau_R \sim 1/kA$. This indicates that the transition occurs according to the compression rate, and the mixing front is stationary under constant compression rate. For $\tau_D<<\tau_R$, the scaling laws are; $w \propto Pe^{-1/2}$, $R_{max} \propto Da$, $I \propto DaPe^{-1/2}$ and for $\tau_D>>\tau_R$, the scaling laws are; $w \propto Pe^{-1/6}Da^{-1/3}$, $R_{max} \propto Pe^{2/3}Da$, $I \propto Pe^{1/2}Da^{2/3}$.\\
In porous media, we replace the diffusion in the scaling laws by the dispersion. For the case of porous media co-flow, the dispersion coefficient is approximately proportional to velocity $D_{disp}\sim v$ by ignoring the molecular diffusion. When $\tau_D<<\tau_R$ in early time regime, we replace the diffusion in $w \propto D^{1/2}t^{1/2}$ by dispersion as $w \propto D_{disp}^{1/2}t^{1/2} \sim v^{1/2}t^{1/2}$. The maximum reaction rate is independent of dispersion, as $R_{max} : const$ in the same reasoning as in the diffusion case. The reaction intensity is thus $I \sim wR_{max} \sim v^{1/2}t^{1/2}$. When $\tau_D>>\tau_R$ in later time regime, we have the same scaling laws over time as in Hele-Shaw case over time because the dispersion is constant. The prefactors of these scaling laws over time at later time regime is determined by the values at the transition. Since the transition time $t_c \sim 1/kA$ is independent of dispersion, the scaling laws in the later time regime includes the velocity dependency in the same form as in the early time regime $w \propto v^{1/2}t^{1/6}$, $R_{max} \propto t^{-2/3}$, $I \propto v^{1/2}t^{-1/2}$; \\
We finally consider the case of porous media saddle-flow. The scaling laws should again depend on $\tau_D=s^2/D_{disp}$ and $\tau_R= 1/kA$. Around the stagnation point, our previous study showed that the mixing zone of the conservative tracer scaled weaker than $\sqrt{x}$ due to small velocity, whereas the mixing zone scales as $\sqrt{x}$ far from the stagnation point. Since the fluid is accelerating over x, we have $D_{disp}\propto \alpha\gamma x$. This makes $\tau_D$ keep decreasing over x close to the stagnation point. When the scaling of $s$ becomes $s\propto\sqrt{x}$, $\tau_D$ becomes constant over distance. If the dispersion dominates reaction $\tau_D<<\tau_R$, we may apply the scaling laws of early time regime in reaction-diffusion system. The width of the reaction zone is determined by the size of the mixing zone as $w\propto\sqrt{x}$, $R_{max};Const.$ and $I\propto x^{1/2}$.\\
We summarized the scaling laws in porous media in Table.\ref{tableScalings}. The scaling laws of porous media co-flow and saddle-flow were consistent with the simulations assuming the pore scale complete mixing. It would be useful to have the theoretical derivation of $\tau_D>>\tau_R$ in saddle-flow and velocity dependency of saddle-flow in a future study.\\

\begin{table*}
\centering
\caption{\label{tableParams}Theoretically expected scaling laws for each scenario in porous media when the pore scale mixing does not influence the overall scaling laws.}
\begin{tabular}{ccccccc}
\hline
 &\multicolumn{2}{c}{Width}&\multicolumn{2}{c}{Maximum reaction rate}&\multicolumn{2}{c}{Intensity}\\
&$\tau_D<<\tau_R$&$\tau_D>>\tau_R$&$\tau_D<<\tau_R$&$\tau_D>>\tau_R$&$\tau_D<<\tau_R$&$\tau_D>>\tau_R$\\ \hline
 Co-flow, time&$t^{1/2}$&$t^{1/6}$ &Const.&$t^{-2/3}$& $t^{1/2}$&$t^{-1/2}$ \\
  Co-flow, velocity&$v^{1/2}$&$v^{1/2}$ &Const.&$t^{-2/3}$& $v^{1/2}$&$v^{1/2}$ \\
 Saddle-flow, distance &$x^{1/2}$&N.A.&Const.&N.A.&$x^{1/2}$&N.A.\\
  Saddle-flow, velocity &N.A.&N.A.&N.A.&N.A.&N.A.&N.A.\\
 \hline
\end{tabular}
\label{tableScalings}
\end{table*}

\section{Results of conservative tracer experiments}
Here we show the results of conservative tracer experiments, highlighting that the mixing width is independent of Pe. For the co-flow case without porous media, in Lagrangian framework, the width of the mixing zone of the conservative tracer $w_c$ develops as $w_c \propto \sqrt{Dt}$, where D is diffusion coefficient and $t = x/v_x$, where $v_x$ is the uniform fluid velocity. In the porous media case, we replace the diffusion coefficient by dispersion coefficient $D_{disp}=D+\alpha v_x$, where $\alpha$ is dispersivity and v is velocity. When the velocity is large enough, $D_{disp}\sim \alpha v_x$. From $t=x/v_x$, we have $w_c\propto \sqrt{\alpha x}$, which is independent of velocity. In case of saddle-flow in Hele-Shaw cell, lamella description of mixing theory gives the relationship $w_c\propto \sqrt{D/\gamma}$, where $\gamma=v_x/x$ is the compression rate \citep{Villermaux2019}. In case of porous media, by replacing diffusion coefficient by dispersion coefficient, we end up with $w_c\propto \sqrt{\alpha x}$, which is 
independent of velocity. This is supported by the experimental results. More detailed discussion about the conservative tracer experiments is available in our previous study.

\begin{figure}
    \centering
    \includegraphics[width=\textwidth,height=\textheight,keepaspectratio]{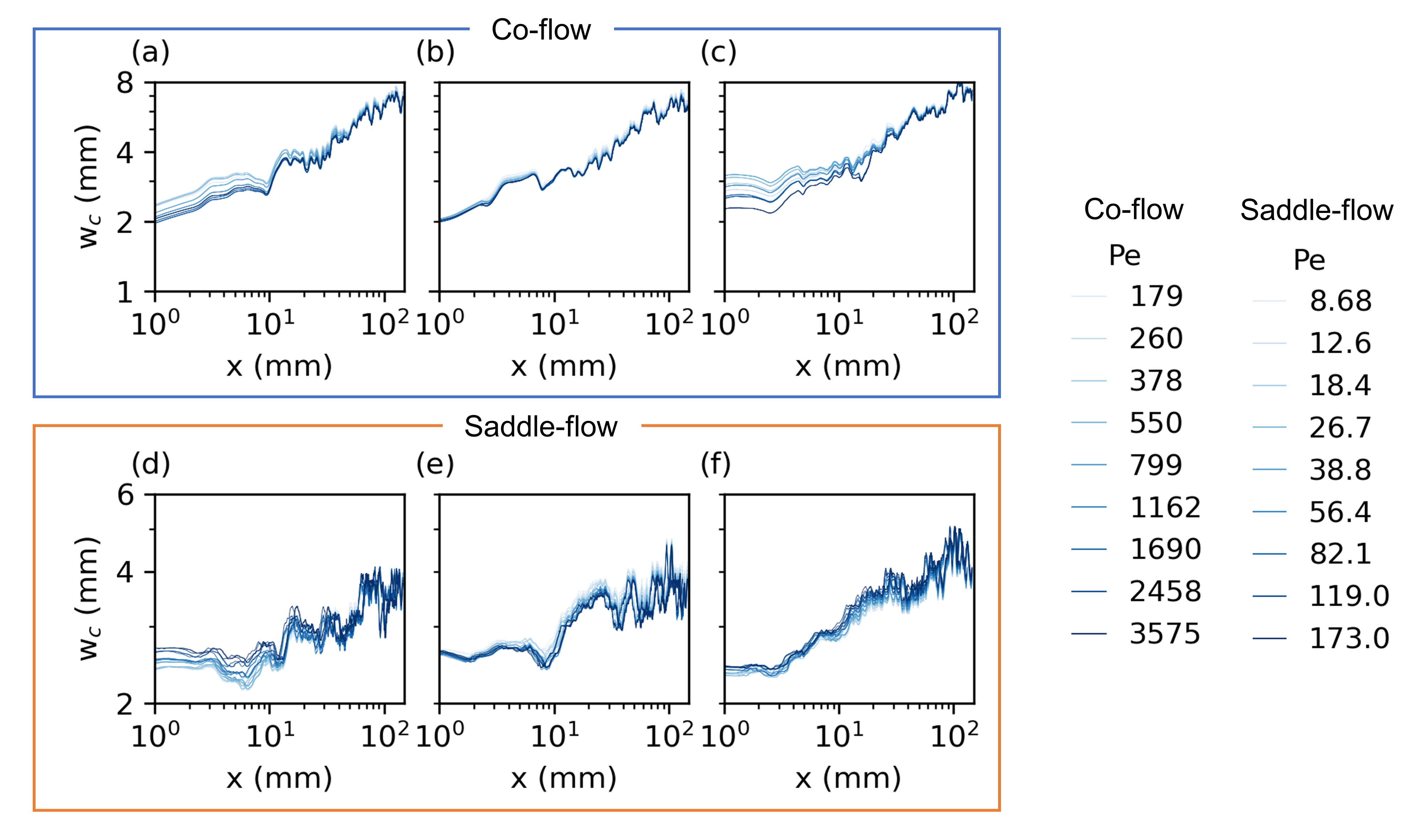}
    \caption{The results of conservative tracer experiments. The triplicated experiments of co-flow (a)-(c) and the triplicated experiments of saddle flow (d)-(f).}
    \label{figCons}
\end{figure}

\bibliographystyle{jfm}
%\bibliography{references}

\end{document}